\documentclass[aps,pra,twocolumn,showpacs]{revtex4}

\usepackage{graphicx}

\newcommand{\cmax}{\ensuremath{\cos_{\mathrm{max}}}}
\newcommand{\pmax}{\ensuremath{P_{\mathrm{max}}}}

\newcommand{\spmax}{\ensuremath{s_i(\pmax)}}

\begin{document}

\title{Optimally Controlled Field-Free Orientation of the Kicked Molecule}

\author{Claude M. Dion}
\affiliation{Department of Physics, Ume{\aa} University, SE-901\,87 Ume{\aa},
Sweden}

\author{Arne Keller}
\author{Osman Atabek}
\affiliation{Laboratoire de Photophysique Mol\'{e}culaire du CNRS,
B\^{a}timent 210, Campus d'Orsay, F-91405 Orsay, France}

\date{\today} 

\begin{abstract}
  Efficient and long-lived field-free molecular orientation is
  achieved using only two kicks appropriately delayed in time.  The
  understanding of the mechanism rests upon a molecular target state
  providing the best efficiency versus persistence compromise.  An
  optimal control scheme is referred to for fixing the free parameters
  (amplitudes and the time delay between them).  The limited number of
  kicks, the robustness and the transposability to different molecular
  systems advocate in favor of the process, when considering its
  experimental feasibility.
\end{abstract}


\maketitle

\section{Introduction}

Molecular orientation, in particular during the free evolution of the
system, has recently been abundantly discussed in the literature as a
process playing an important part in a variety of laser-molecule
control issues, among which are chemical
reactivity~\cite{orient:brooks76}, nanoscale
design~\cite{focus:seideman97b,manip:dey00}, surface
processing~\cite{orient:tenner91,focus:mcclelland93}, and attosecond
time scale pulse production~\cite{align:bandrauk03,align:denalda04}.
A basic mechanism by which orientation is achieved involves sudden
optical excitation, such as half-cycle pulses (HCP), that impart a
kick to the molecule which orients itself along the polarization
vector of the linearly-polarized electromagnetic field (the kick
mechanism)~\cite{hcp:dion01,orient:machholm01,orient:matos-abiague03}.

Application of a series of kicks is in fact a general control strategy
that can enhance molecular
alignment~\cite{align:leibscher03,align:leibscher04} or even lead to a
squeezing of atoms in an optical
lattice~\cite{lat:leibscher02,lat:windell02}.  Moreover, alignment by
a pair of pulses has been experimentally
achieved~\cite{align:lee04,align:bisgaard04}.  The generic strategy,
as first suggested in Ref.~\cite{control:averbukh01}, rests on the
application of a sudden impulse when a certain observable reaches a
maximum (or minimum), such as the expectation value of $\cos \theta$
(which measures orientation, $\theta$ being the angle positioning the
molecular axis with respect to the laser polarization) for the control
of orientation.

In this paper we present a different approach based on a \emph{target
  state}~\cite{align:sugny04a}, instead of an observable, which allows
to consider not only the efficiency of the orientation ($\left\langle
  \cos \theta \right\rangle$), but also its persistence.  This stems
from the fact that, when dealing with field-free orientation of a
molecule, a compromise between efficiency and duration has to be
looked for in the optimization criterion~\cite{orient:benhajyedder02}.
Although the generic strategy (i.e., kicking when an observable
reaches its maximum value) can be applied also to reach the target
state~\cite{control:sugny05a}, we show here, that this strategy is far
from optimal, and that two kicks can be sufficient to come within 1\%
of the target state.

\section{Method}

\subsection{Model}
\label{sec:model}

A diatomic molecule illuminated by a moderate amplitude HCP is
described within a model of a rigid rotor interacting with the field
through its permanent dipole (polarizability interaction neglected).
The time-dependent Schr\"{o}dinger equation (TDSE) accounting for the time
evolution of the system is (in atomic units)
\begin{equation}
i \frac{d}{dt} \psi(\theta, \varphi; t) = \left[ B \hat{J}^2 -\vec{\mu}_{0}
  \cdot \vec{\epsilon} \mathcal{E}(t) \right] \psi(\theta, \varphi;
  t),
\label{eq:tdse}
\end{equation}
where the molecule is described by its rotational constant $B$ and
permanent dipole moment $\vec{\mu}_{0}$, and the HCP is characterized
by its amplitude $\mathcal{E}(t)$ and unit polarization vector
$\vec{\epsilon}$.  $\hat{J}$ is the angular momentum operator.  The
wave function involves two angular variables: the polar $\theta$ and
azimuthal $\varphi$ angles.  Due to cylindrical symmetry around
$\vec{\epsilon}$, the motion associated with the azimuthal angle
$\varphi$ is completely separated and is not considered hereafter. As
a consequence, the projection quantum number $M$ of $\vec{J}$ on
$\vec{\epsilon}$ is fixed.  The polar angle $\theta$ takes part (apart
from the analytical characterization of $\hat{J}^2$) in the dot
product $\vec{\mu}_{0} \cdot \vec{\epsilon}$ in Eq.~(\ref{eq:tdse}),
through its cosine.  The short duration $\tau$ of the HCP, as compared
to the molecular rotational period $T_\mathrm{rot} = \pi/B$, is
accounted for through a dimensionless small parameter $\varepsilon =
\tau / T_\mathrm{rot} = \tau B/\pi$ less than one.  A time scaling is
achieved~\cite{tdupt:daems03,align:sugny04b} by introducing a variable
$s = t/T_\mathrm{rot}$ ($s \in [0,\varepsilon]$ during the HCP pulse),
leading to the following form of the TDSE~(\ref{eq:tdse})
\begin{equation}
i \frac{d}{ds} \psi(\theta, \varphi; s) = \left[ \pi \hat{J}^2
  -E(s) \cos\theta \right] \psi(\theta, \varphi; s),
\label{eq:tdupt}
\end{equation}
where $E(s) = \pi \frac{\mu_0}{B} \mathcal{E}(T_\mathrm{rot} s)$.
A sudden approximation, where the small perturbation parameter is the
duration of the pulse $\varepsilon$, provides the wave packet at time
$s$, as a result of successive applications of two unitary evolution
operators on the initial state $\psi(\theta, \varphi; s=0)$ (taken as
a pure quantum state), 
\begin{equation}
\psi(\theta, \varphi; s) \simeq U_0(s) U_{\mathrm{a}} \psi(\theta,
\varphi; s=0),
\label{eq:prop}
\end{equation}
where
\begin{equation}
 U_0(s) = \exp \left[ -i \pi \hat{J}^2 s \right]
\end{equation}
\label{eq:U0}
and
\begin{equation}
 U_{\mathrm{a}} = \exp \left[ i A \cos \theta \right],
\label{eq:Ua}
\end{equation}
with $A = \int_{0}^{\varepsilon} E(s) ds$ a dimensionless parameter
combining molecule and field characteristics, integrated over the
whole pulse duration.  The initial molecular pure quantum state is
taken as the isotropic spherical harmonic $Y_{J=M=0}
(\theta,\varphi)$.  The dynamical picture, which emerges from
Eqs.~(\ref{eq:prop})--(\ref{eq:Ua}) for a single HCP, is that a
unitary operator $U_{\mathrm{a}}$ imparts a kick, measured in terms of
a strength $A$ times $\cos\theta$, to the molecule in its initial
state.  This produces a rotational excitation bringing the molecule
from a completely isotropic angular distribution to a more oriented
configuration.  The further field-free evolution is monitored by
$U_0(s)$.  In the case where a series of HCPs is considered, the
evolution operators $U_0(s) U_{\mathrm{a}}$ have to be applied for
each individual pulse, taking into account the time delays between
each.  It is also worthwhile noting that introducing the dimensionless
parameters $\varepsilon$, $A$ and variable $s$ helps to free oneself
from a specific molecule-plus-field system.  Actually, molecular and
field characteristics are combined in such a way that a large
rotational constant $B$ can be overcome by a shorter pulse duration
$\tau$, the relevant condition being $\varepsilon \ll 1$.  Similarly,
a small permanent dipole $\mu_0$ can be overcome by a stronger field
amplitude $\mathcal{E}$, the relevant parameter $A$ being proportional
to the product $\mu_0 \mathcal{E}/B$.  As for the rescaling of time,
it gives access to results in terms of a dimensionless time $s$ that
is taken as a fraction of the rotational period, which again is
molecule independent.

\subsection{Target state}
\label{sec:target}

A measure of orientation is given by the dynamical expectation value
of  $\cos \theta$, i.e.,
\begin{equation}
\left\langle \cos \theta \right\rangle (s) = \left\langle \psi
  (\theta, \varphi; s) \right| \cos\theta \left| \psi
  (\theta, \varphi; s) \right\rangle,
\end{equation}
the optimization of the orientation being in relation with the
maximization (or minimization) of $\left\langle \cos \theta
\right\rangle (s)$.  We have recently developed a generic strategy
that, when applied to $\cos \theta$ taken as an operator, can be
summarized as following~\cite{align:sugny04a}:

(i) The physical Hilbert space in which $\cos \theta$ is acting is
reduced to a finite subspace $\mathcal{H}^{(N)}$ of dimension $N$.
The expectation value of the projection of $\cos \theta$ on this
subspace is time periodic and can be represented, in the basis of
spherical harmonics, as a finite $N\times N$-dimensional matrix with
discretized, bounded eigenvectors (as opposed to $\cos \theta$ which
has a continuous spectrum).

(ii) A target state can be defined as the eigenstate of the projection
of $\cos \theta$ on $\mathcal{H}^{(N)}$ corresponding to its highest
(or lowest) eigenvalue.  Such a state can be explicitly calculated, in
the basis of spherical harmonics, by diagonalizing the corresponding
matrix.  The full advantage of the dimensionality reduction remains in
the fact that a smaller $N$ involves a lower rotational excitation
that allows for a longer duration of the orientation after the pulse
is over (see Fig.~1 of Ref.~\cite{align:sugny04a}).  The most exciting
observation is that a target state calculated within a subspace as
small (with respect to dimensionality) as $N=5$ already leads to an
excellent orientation efficiency of $\left\langle \cos \theta
\right\rangle \sim 0.9$.  In other words, the target state fulfills
the two requirements of the orientation control problem, and in that
respect is far superior to all other intuitive criteria that have been
previously used~\cite{orient:benhajyedder02}.

(iii) A strategy is proposed to reach this target state by applying a
series of identical short pulses at times when $\left\langle \cos
\theta \right\rangle$ reaches its maximum (or minimum) following
field-free evolution.  Consequently, the corresponding wave function
converges to the target state.  The robustness of the strategy has
been checked against the pulse strength and the time delays.  For
completeness, we have also to mention a similar strategy (leading to
similar results) that consists in applying the pulses every time the
projection of the time-evolved wave packet reaches its maximal
projection on the target state.

From this theory, that can actually be generically transposed to other
control issues, the recipes that emerge for a possible control of
molecular orientation is the application of a train of short and
identical pulses with a given total area (within 10 to 15\% of
accuracy) and respecting predetermined time delays between successive
pulses (within 10\% of accuracy).  For typical cases, 10 to 20 pulses
with $A=1$ are necessary to reach the target state.  But obviously
this strategy taken as a whole (with the values of the time delays,
the number of pulses and their integrated amplitude $A$) is not
unique.  However, it indicates that a train of short, time-delayed
pulses induces a repeated kick mechanism at specific molecular
response times that improves the efficiency and the duration of
orientation.  It is precisely this information that serves here as a
basis for a numerical optimal control scheme conducted using an
evolutionary strategy (ES)~\cite{ga:michalewicz96}. 

The target state being clearly identified, the optimization aims to
maximize the projection of the instantaneous wave packet on the
target.  More precisely, the wave packet at time $s$ being expanded on
the basis of spherical harmonics $Y_{J,M=0} (\theta, \varphi)$ as
\begin{equation}
\psi (\theta, \varphi; s) = \sum_{J=0}^{\infty} c_J(s) Y_{J,M=0} (\theta,
\varphi), 
\end{equation}
the probability to be maximized is 
\begin{equation}
P(s) = \left| \mathbf{c}^{*}(s) \cdot \mathbf{c}_{\mathrm{target}}
\right|^{2},
\end{equation}
where $\mathbf{c}^{*}(s) \equiv \left( c_0^{*}(s), c_1^{*}(s), \ldots,
\right)$ and $\mathbf{c}_{\mathrm{target}}$ is the corresponding
column vector of the weighting coefficients of the target state on the
same basis of spherical harmonics.  The ES deal as usual with the
minimization of a criterion defined here as
\begin{equation}
j \equiv 1 - P(\spmax),
\label{eq:crit}
\end{equation}
where $\spmax$ is the time for which $P$ reaches a maximum during the
free evolution over a rotational period following the radiative
interaction.  The parameters of the optimization are the amplitudes ($A$)
and time intervals characterizing a train of kicks that can be
produced by HCPs.  The ES is implemented using the Evolving Objects
library (EOlib)~\cite{eo:keijzer02,EOlib}.

\section{Results}

All the calculations that are presented here deal with a target state
in a $N = 5$-dimensional finite subspace $\mathcal{H}^{(5)}$(the
maximum allowed rotational excitation $J_\mathrm{max} = 4$ being the
one previously retained as satisfying the best post-pulse orientation
efficiency/duration compromise~\cite{align:sugny04a}).  The wave
packet by itself is propagated in a larger, although finite, subspace
and thus may reach higher rotational states. We note that the use of a
target state defined in a reduced Hilbert subspace dispenses from
having to penalize with respect to the total pulse intensity.  Too
strong pulses will necessarily move the system outside the subspace
and thus reduce the value of the projection on the target state.
Nevertheless, using a target state is not equivalent to penalizing
since, for the same pulse energy, there might exist states that show
better orientation but with less persistence.  Only the target state
ensures that the efficiency/persistence compromise is achieved.  The
optimal control strategy is guided by two different approaches,
depending upon the parameter space chosen.

\subsection{Time delays as the only parameters}

For a given number of pulses, the only task conferred to the ES
algorithm is the optimal determination of the time delays between the
pulses for a minimization of $j$ [Eq.~(\ref{eq:crit})].  The pulses
are considered identical, with all other parameters taken such that $A
= 1$.  The results for 3 and 4 pulses are displayed in
tables~\ref{tab:3k} and \ref{tab:4k}, collecting the values of the
time delays $\delta_i$ between pulses $i$ and $i+1$ and their times of
application $s_i$, together with the values reached for $P(\spmax)$
after each successive kick.  Are also given the time intervals
$\delta_i(\cmax)$ and $\delta_i(\pmax)$ between the $i$th kick and the
next maxima of $\left\langle \cos \theta \right\rangle (s)$ and of
$P(s)$, respectively, for comparison with the strategies of
Ref.~\cite{align:sugny04a}.
\begin{table}
\begin{center}
\begin{ruledtabular}
\begin{tabular}{|c|l|l|l|l|l|} 
$i$ & $\delta_i$ & $s_i$ & $\delta_i(\cmax)$ & $\delta_i(\pmax)$ & $P_{\mathrm{max}}$
\\ \hline 
1 &        & 0.     &        &        & 0.3913 \\
  & 0.1065 &        & 0.2070 & 0.1607 &  \\
2 &        & 0.1065 &        &        & 0.6685 \\ 
  & 0.0029 &        & 0.0991 & 0.0855 &  \\ 
3 &        & 0.1094 &        &        & 0.8905 
\end{tabular}
\end{ruledtabular}
\end{center}
\caption{Optimal delays $\delta_i$ (and corresponding times between
  $s_i$) between 3 kicks of constant amplitude $A = 1$, along with the
  projection $P_{\mathrm{max}}$ on the target state.  $\delta_i(\cmax)$ and $\delta_i(\pmax)$ are the time intervals between the $i$th kick and the
next maxima of $\left\langle \cos \theta \right\rangle (s)$ and of
$P(s)$, respectively.}
\label{tab:3k}
\end{table}
\begin{table}
\begin{center}
\begin{ruledtabular}
\begin{tabular}{|c|l|l|l|l|l|} 
$i$ & $\delta_i$ & $s_i$ & $\delta_i(\cmax)$ & $\delta_i(\pmax)$ & $P_{\mathrm{max}}$
\\ \hline 
1 &                     & 0.     &        &        & 0.3913  \\
  & 0.2388              &        & 0.2070 & 0.1607 &   \\
2 &                     & 0.2388 &        &        & 0.4848  \\
  & 0.9901              &        & 0.0907 & 0.0962 &   \\
3 &                     & 1.2289 &        &        & 0.8018  \\
  & $1.\times 10^{-12}$ &        & 0.0884 & 0.0853 &  \\
4 &                     & 1.2289 &        &        & 0.9787
\end{tabular}
\end{ruledtabular}
\end{center}
\caption{Same as for Tab.~\ref{tab:3k}, but with four identical pulses.}
\label{tab:4k}
\end{table}
Concerning the three-pulse model of
Tab.~\ref{tab:3k}, two observations can be made, showing that the
general theory of Ref.~\cite{align:sugny04a}, as summarized 
in Sec.~\ref{sec:target},
is neither unique nor optimal.  A value of $P=0.8905$ is reached using
the optimal time delays $\delta_i$, far better than the one that can
be reached by applying the kicks precisely at the maxima of
$\left\langle \cos \theta \right\rangle (s)$, yielding $P = 0.5277$.
The comparison between $\delta_i$'s and $\delta_i(\cmax)$ and
$\delta_i(\pmax)$'s shows that the optimal strategy is to apply the
pulses before the maxima of $\left\langle \cos \theta \right\rangle
(s)$ or of $P(s)$.  The situation is different for the 4-pulse model,
Tab.~\ref{tab:4k}, advocating again for the non-uniqueness of the
solution.  The second pulse in particular is applied after a time
delay $\delta_1$ larger than $\delta_1(\cmax)$ or $\delta_1(\pmax)$.
An excellent value for $P$ is obtained, showing that after 4 pulses
the molecular state that is reached is close to the target within 2\%.
But, even more interestingly, the third and fourth pulses are applied
at times very close to one rotational period (i.e., $\delta_2 = 0.99$,
$\delta_3 = 0.00$) after the second kick.  Due to the periodicity,
this amounts to applying simultaneously three identical pulses after a
time delay $\delta_1$ corresponding to the second pulse.  Still
another way of analyzing the situation consists in applying a first
pulse with an amplitude $A_1 = 1.0$ and after a time delay of
$\delta_1 = 0.2388$ applying a second pulse of amplitude $A_2 = 3.0$.
Such a strategy has actually been checked and leads to a final
projection $P = 0.9776$, very close to the one displayed in
Tab.~\ref{tab:4k}, i.e., $P = 0.9787$.

We note that a similar result is also obtained with five kicks,
allowing then to reach $P=0.9928$.  Better results are not reached
with more kicks as the subsequent ones have the counter effect of
increasing the rotational excitation, and thus of pushing the system
outside the Hilbert space $\mathcal{H}^{(5)}$ where the target
resides.

\subsection{Time delays and amplitudes as parameters}

Referring again to the strategy of implementing in the ES what is
learned from previous attempts, we extend the parameter space such as
to account for the variation of both the time delays $\delta_i$ and
the amplitudes $A_i$.  In addition, we restrict the optimization
scheme to a 2-pulse model, which results into the rather simplified
task for the ES of providing merely with 3 parameters: $A_1$, $A_2$,
and $\delta_1$.

\begin{figure}
\centerline{\includegraphics[width=0.5\textwidth]{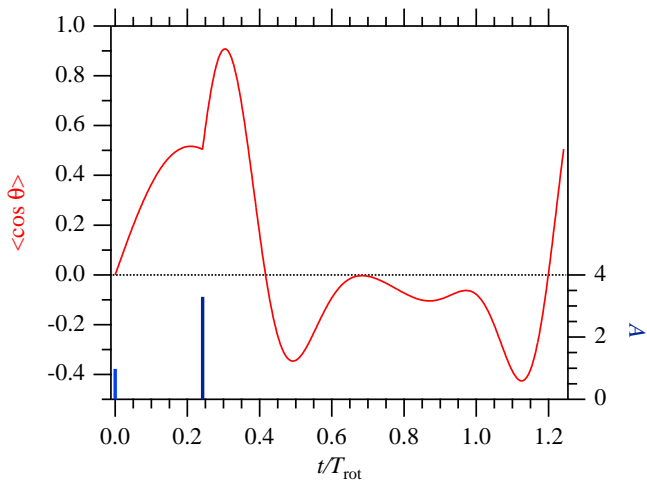}}
\caption{Resulting orientation, as expressed by $\left\langle \cos
    \theta \right\rangle$, for the optimal 2-kick solution, along with
  the amplitudes of the kicks.}
\label{fig:cosavg}
\end{figure}

\begin{figure}
\centerline{\includegraphics[width=0.5\textwidth]{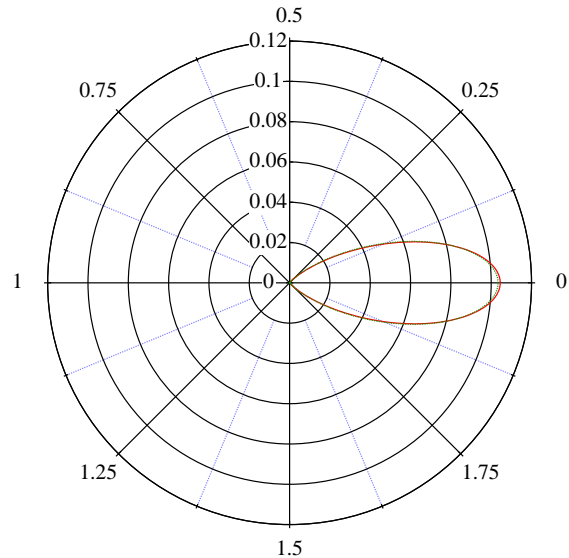}}
\caption{Polar plot of the angular
  distribution reached at the maximum of orientation, $t = 0.3042
  T_\mathrm{rot}$ (solid line) compared with that of the target state
  (dotted line).}
\label{fig:polar}
\end{figure}

The optimization then yields the strategy of giving a first kick of
amplitude $A_1 = 0.9741$, followed by the second of amplitude $A_2 =
3.2930$ after a delay $\delta_1 = 0.2419$, allowing to reach $P =
0.9886$.  This confirms the previous observations: the target state is
easily reached using only two pulses and the intensity of the kicks
must be restrained so that highly excited rotational states are not
populated.  The resulting time evolution of $\left\langle \cos \theta
\right\rangle$ is given in Fig.~\ref{fig:cosavg}.  A value of
$\left\langle \cos \theta \right\rangle = 0.9078$ is reached, slightly
greater than that of the target (0.9062).  The corresponding angular
distribution at maximum orientation is shown in Fig.~\ref{fig:polar},
along with the target state: the two are virtually indistinguishable
at this scale.

We have checked the robustness of this strategy by varying the
parameters by 10\% of their optimal value.  The smallest value of $P$
obtained is then 0.9689, so the results remains quite
close to the target.  Orientation efficiency remains within 2.2\%,
whereas its duration is shorten by at most 9\%.

\section{Conclusion}

In conclusion, using evolutionary strategies, we found the optimal way
to kick a molecule with short HCPs in order to reach a target state
corresponding to an oriented molecule.  The solution turns out to be
very efficient, allowing to reach the target state (within 1\%) with
only two pulses, instead of the approximately 15 pulses of the
previous mathematically built (but not unique) strategy of
Ref.~\cite{align:sugny04a} for the same system.  This advocates for
great experimental feasibility and, even more importantly, it points
out the broad interest of the overall methodology.  As has been
already shown, the mathematically clear depiction of a quantum target
state in a finite dimensional subspace can be conducted for a large
class of observables (some examples in comparing systems interacting
with a thermal baths or other dissipative environments are provided in
Ref.~\cite{control:sugny05b}).  Once the target state is defined, ES
can be successfully run with a simple criterion of maximum projection
on the target state referring as a basic mechanism to a train of
kicks.  The only parameters to be optimized are the time delays and
amplitudes.  A small number of such kicks, reachable within modest
experimental environment, allow for a remarkably efficient and
persistent control.

\begin{acknowledgments}
The authors thank Dr.\ Anne Auger for her help in setting up the
evolutionary strategies.
\end{acknowledgments}

\end{document}